\documentclass[sigconf]{acmart}
\usepackage[utf8]{inputenc}

\usepackage{algorithm}
\usepackage{algpseudocode}
\usepackage{bm}
\usepackage{centernot}
\usepackage{doi}
\usepackage{listings}
\usepackage{siunitx}
\usepackage{tikzit}
\usepackage{tikz}
\usepackage{xspace}

\newcommand{\SWH}{Software Heritage\xspace}
\newcommand{\SWHGD}{\SWH Graph Dataset\xspace}
\newcommand{\TODO}[1]{\textcolor{red}{\textbf{TODO:} \footnotesize #1}}

\newcommand{\RQBOX}[1]{\noindent\fbox{\parbox{0.97\linewidth}{\bfseries #1}}}
\newcommand{\figref}[1]{\figurename~\ref{#1}}

\newcommand{\forked}{\ensuremath{\rightsquigarrow}}
\newcommand{\forkedI}{\ensuremath{\rightsquigarrow_1}}
\newcommand{\forkedII}{\ensuremath{\rightsquigarrow_2}}
\newcommand{\forkedIII}{\ensuremath{\rightsquigarrow_3}}

\newcommand{\notforkedI}{\ensuremath{\not\rightsquigarrow_1}}

\newcommand{\network}{\ensuremath{\mathcal{N}}}
\newcommand{\networkI}{\ensuremath{\mathcal{N}^1}}
\newcommand{\networkII}{\ensuremath{\mathcal{N}^2}}
\newcommand{\networkIII}{\ensuremath{\mathcal{N}^3}}
\newcommand{\clique}{\ensuremath{\mathcal{C}}}

\colorlet{punct}{red!60!black}
\definecolor{background}{HTML}{EEEEEE}
\definecolor{delim}{RGB}{20,105,176}
\colorlet{numb}{magenta!60!black}

\lstdefinestyle{sql}{
    language=SQL,
    basicstyle=\footnotesize\ttfamily,
    showstringspaces=false,
    breaklines=true,
    frame=lines,
    backgroundcolor=\color{background},
}

\lstdefinelanguage{json}{
    basicstyle=\footnotesize\ttfamily,
    showstringspaces=false,
    breaklines=true,
    frame=lines,
    backgroundcolor=\color{background},
    literate=
     *{0}{{{\color{numb}0}}}{1}
      {1}{{{\color{numb}1}}}{1}
      {2}{{{\color{numb}2}}}{1}
      {3}{{{\color{numb}3}}}{1}
      {4}{{{\color{numb}4}}}{1}
      {5}{{{\color{numb}5}}}{1}
      {6}{{{\color{numb}6}}}{1}
      {7}{{{\color{numb}7}}}{1}
      {8}{{{\color{numb}8}}}{1}
      {9}{{{\color{numb}9}}}{1}
      {:}{{{\color{punct}{:}}}}{1}
      {,}{{{\color{punct}{,}}}}{1}
      {\{}{{{\color{delim}{\{}}}}{1}
      {\}}{{{\color{delim}{\}}}}}{1}
      {[}{{{\color{delim}{[}}}}{1}
      {]}{{{\color{delim}{]}}}}{1},
}

\makeatletter
\pgfkeys{/tikz/.cd,
  tab height/.initial=3pt,
  tab width/.initial=10pt,
  tab slope/.initial=1.5pt,
  cover xoff/.initial=5pt,
  cover yoff/.initial=2pt,
}
\newlength\pgf@ytt
\newlength\pgf@xe
\newlength\pgf@xf
\newlength\pgf@xg
\newlength\pgf@yg
\newlength\pgf@xh
\newlength\pgf@xo  

\pgfdeclareshape{document}{ \inheritsavedanchors[from=rectangle]
  \inheritanchorborder[from=rectangle]
  \inheritanchor[from=rectangle]{center}
  \inheritanchor[from=rectangle]{north}
  \inheritanchor[from=rectangle]{south}

  \savedmacro\tabheight{\edef\tabheight{\pgfkeysvalueof{/tikz/tab height}}}
  \savedmacro\tabwidth{\edef\tabwidth{\pgfkeysvalueof{/tikz/tab width}}}
  \savedmacro\tabslope{\edef\tabslope{\pgfkeysvalueof{/tikz/tab slope}}}
  \savedmacro\coverxoff{\edef\coverxoff{\pgfkeysvalueof{/tikz/cover xoff}}}
  \savedmacro\coveryoff{\edef\coveryoff{\pgfkeysvalueof{/tikz/cover yoff}}}

\anchor{west}{
    \pgf@process{\northeast}\pgf@ya=.5\pgf@y \pgf@process{\southwest}\pgf@y=.5\pgf@y }
\anchor{east}{\pgf@process{\southwest}\pgf@ya=.5\pgf@y \pgf@process{\northeast}\pgf@y=.5\pgf@y }
  \backgroundpath{\southwest \pgf@xa=\pgf@x \pgf@ya=\pgf@y
    \northeast \pgf@xb=\pgf@x \pgf@yb=\pgf@y
\setlength{\pgf@ytt}{\pgf@yb}
    \advance\pgf@ytt by \tabheight
    \setlength{\pgf@xe}{\pgf@xa}
    \advance\pgf@xe by \tabwidth \setlength{\pgf@xf}{\pgf@xe}
    \advance\pgf@xf by \tabslope \setlength{\pgf@xg}{\pgf@xa}
    \advance\pgf@xg by \coverxoff \setlength{\pgf@xh}{\pgf@xb}
    \advance\pgf@xh by \coverxoff
    \setlength{\pgf@yg}{\pgf@yb}
    \advance\pgf@yg by-\coveryoff \pgfpathmoveto{\pgfpoint{\pgf@xa}{\pgf@ya}}
    \pgfpathlineto{\pgfpoint{\pgf@xa}{\pgf@ytt}}
    \pgfpathlineto{\pgfpoint{\pgf@xe}{\pgf@ytt}}
    \pgfpathlineto{\pgfpoint{\pgf@xf}{\pgf@yb}}
    \pgfpathlineto{\pgfpoint{\pgf@xb}{\pgf@yb}}
    \pgfpathlineto{\pgfpoint{\pgf@xb}{\pgf@ya}} \pgfpathmoveto{\pgfpoint{\pgf@xb}{\pgf@ya}}
\pgfpathclose
\pgfpathmoveto{\pgfpoint{\pgf@xa}{\pgf@ya}}
\pgfpathlineto{\pgfpoint{\pgf@xb}{\pgf@ya}}
    \pgfpathclose
  }
}
\makeatother

\tikzstyle{origin}=[fill=white, draw=black, shape=cloud, tikzit fill={rgb,255: red,0; green,166; blue,255}]
\tikzstyle{revision}=[fill={rgb,255: red,230; green,242; blue,242}, draw=black, shape=regular polygon, regular polygon sides=3, minimum width=20pt, inner sep=0pt]
\tikzstyle{directory}=[fill={rgb,255: red,218; green,219; blue,231}, draw=black, shape=document, tab width=6pt, tab height=2pt, minimum width=14pt, minimum height=8pt]
\tikzstyle{content}=[fill={rgb,255: red,198; green,192; blue,193}, draw=black, shape=cylinder, shape border rotate=90, minimum width=12pt, minimum height=13pt]
\tikzstyle{release}=[fill={rgb,255: red,255; green,252; blue,204}, draw=black, shape=star]
\tikzstyle{snapshot}=[fill=white, draw=black, shape=rectangle, minimum width=15, minimum height=9]

\tikzstyle{arrow}=[->, thick]
\tikzstyle{cluster}=[-, draw=gray, smooth, dashed]
\tikzstyle{red arrow}=[draw=red, ->, thick]

\setcopyright{licensedothergov}
\acmConference[MSR '20]{17th International Conference on Mining Software Repositories}{October 5--6, 2020}{Seoul, Republic of Korea}
\acmBooktitle{17th International Conference on Mining Software Repositories (MSR '20), October 5--6, 2020, Seoul, Republic of Korea}
\acmPrice{15.00}
\acmDOI{10.1145/3379597.3387450}
\acmISBN{978-1-4503-7517-7/20/05}

\acmConference[MSR 2020]{Mining Software Repositories 2020}{25-26 May, 2020}{Seoul, South Korea}

\title[Forking Without Clicking: How to Identify Forks]{Forking Without Clicking:\\
  on How to Identify Software Repository Forks}
\author{Antoine Pietri}
\email{antoine.pietri@inria.fr}
\orcid{0000-0003-4052-4469}
\affiliation{\institution{Inria}
  \city{Paris}
  \country{France}
}

\author{Guillaume Rousseau}
\email{guillaume.rousseau@u-paris.fr}
\orcid{0000-0003-3583-995X}
\affiliation{\institution{Université de Paris and Inria}
  \city{Paris}
  \country{France}
}

\author{Stefano Zacchiroli}
\email{zack@irif.fr}
\orcid{0000-0002-4576-136X}
\affiliation{\institution{Université de Paris and Inria}
  \city{Paris}
  \country{France}
}

\begin{abstract}
  The notion of \emph{software ``fork''} has been shifting over time from the
  (negative) phenomenon of community disagreements that result in the creation
  of separate development lines and ultimately software products, to the
  (positive) practice of using distributed version control system (VCS)
  repositories to collaboratively improve a single product without stepping on
  each others toes. In both cases the VCS repositories participating in a fork
  share parts of a common development history.

  Studies of software forks generally rely on hosting platform metadata, such
  as GitHub, as the source of truth for what constitutes a fork. These ``forge
  forks'' however can only identify as forks repositories that have been
  created \emph{on} the platform, e.g., by clicking a ``fork'' button on the
  platform user interface. The increased diversity in code hosting platforms
  (e.g., GitLab) and the habits of significant development communities (e.g.,
  the Linux kernel, which is not primarily hosted on any single platform) call
  into question the reliability of trusting code hosting platforms to identify
  forks. Doing so might introduce selection and methodological biases in
  empirical studies.

  In this article we explore various definitions of ``software forks'', trying
  to capture forking workflows that exist in the real world. We quantify the
  differences in how many repositories would be identified as forks on GitHub
  according to the various definitions, confirming that a significant number
  could be overlooked by only considering forge forks. We study the structure
  and size of fork networks, observing how they are affected by the proposed
  definitions and discuss the potential impact on empirical research.
\end{abstract}


\begin{document}
\maketitle

\section{Introduction}
\label{sec:intro}

How developers and software communities work on their projects, and how this
relationship evolves over time, have been topics of interest in software
engineering research for many decades.

Historically, \emph{software ``forking''}~\cite{nyman2016forkhistory} has been
intended as the practice of taking the source code and development history of
an existing software product to create a new, competing product, whose
development will happen elsewhere and taken to different directions. This kind
of ``hard fork'' is enabled by free/open source software (FOSS)
licensing~\cite{fogel2005producingoss} and its possibility is an asset that
guarantees freedom of development; while the actual occurrence of a hard fork
has generally been considered a liability~\cite{robles2012forks} for project
sustainability~\cite{nyman2011-fork-or-not, nyman2014forking-hackers,
  rastogi2016forking}.

In the past decade the rise in popularity of \emph{distributed} version control
systems (DVCS)~\cite{spinellis2005vcs} introduced a significant shift of
paradigm and terminology. The expression ``fork'' is now generally
intended~\cite{zhou2019fork} to refer to the mere technical act of creating a
new VCS repository that contains the full history (at the time of fork) of a
preexisting repository, without an implicit negative connotation (also called
``development forks''~\cite{fogel2005producingoss}). Repository forks can be
created on social coding platforms~\cite{dabbish2012socialcoding,
thung2013network} with as little as a click on a button. Then, while a forked
repository \emph{can} be used to hard fork a project, often it is just a way to
work on software improvements that will be eventually sent back to the
originating project as pull requests~\cite{gousios2014pullrequests} for
integration.

Likely as a consequence of the prevalence of social coding platforms, recent
literature on forks has focused on a single source of truth to determine what
constitutes a fork: metadata provided by code hosting providers, and most
notably GitHub. Clicking the fork button on GitHub indeed, in addition to
cloning development history into a new repository, also registers a ``is forked
from'' relationship between the new repository and its parent. This
relationship forms an ancestry graph that GitHub makes available through its
API and that is what has traditionally been studied as a large, easily
exploitable fork network.

The first drawback of trusting platform metadata as source of truth for what
repository is a fork is that it is platform-specific. One cannot identify as
forks repositories hosted on GitHub that has been forked from, say, GitLab, or
more generally non-GitHub hosted repositories, and vice-versa. Similarly,
although arguably less relevant from a quantitative point of view, one cannot
recognize as forks, say, Git repositories used to collaborate with Subversion
repositories via \texttt{git-svn}. For a fork ecosystem to be properly studied
via the current approach, all the parallel development must happen using the
same VCS and on the same platform. While the prevalence of Git does not seem to
be waning, Git code hosting diversity is increasing, making the
platform-specific part of this problem potentially severe.

A second, more subtle methodological drawback is that trusting platform
metadata introduces a selection bias on both the amount and type of forks that
are considered. The fact that social coding platform strongly encourage, and
sometimes even automate, the creation of forked repositories as the main way to
contribute even the smallest one-liner change, inflates the number of forks.
Many of these (soft) forks will be short-lived in terms of development
activity. Hard forks will comparatively be more long lived and will not
necessarily reside on the same code hosting platform. The example of the Linux
kernel community is revealing in this respect: several copies of the full
development history of Linux exist on GitHub, but are not recognizable as forks
of \texttt{torvalds/linux} according to platform metadata, because kernel
development does not primarily happen on GitHub and kernel developers create
their repositories using \texttt{git clone}.

Fork inflation also results in increased duplication of software artifacts
(source code files or directories, commits, \ldots) across
repositories~\cite{swh-provenance-tr}, which has a significant impact on fork
studies that rely on metrics as simple as repository size (measured as the
number of hosted commits). Filtering out forked repository is a common solution
to this problem, which calls into question \emph{how} to properly identify
forks.

\smallskip

The absence of extensive, homogeneous fork research has been pointed out in the
past as a missing piece~\cite{robles2012forks} in the literature. In this paper
we try to provide methodological tools to enable fork studies that do not
restrict themselves to platform metadata to recognize forks, thereby removing
the constraint of analyzing a single platform and mitigating the risk of
selection biases.

As an alternative to relying on platform metadata to recognize forks we propose
to compare the content of VCS and consider as forks repositories that share
artifacts such as commits or entire source trees. We will explore the impact of
different such definitions and compare their impact in terms of the amount and
structure of forks identified using platform metadata. Specifically, we will
answer the following research questions:

\RQBOX{RQ1: how do code hosting platform information about which VCS
  repositories are forks compare to the presence of shared source code
  artifacts in repositories?}

\RQBOX{RQ2: how are (a) the amount of forks and (b) the structure of fork
  networks affected by fork definitions based on VCS artifact sharing?}

RQ1 will intuitively assess the level of trustworthiness of platform fork
metadata: if many repositories, e.g., share commits but are not identified as
fork by platform metadata, then relying on those metadata alone would appear to
be methodologically dangerous. As one might consider different types of shared
VCS artifacts (commits, source tree directories, individual files, \ldots) as
fork evidence, RQ2 will provide an empirical evaluation of the effects of
basing fork definitions on one or the other.

\paragraph{Paper structure}
Section~\ref{sec:defs} explores the spectrum of fork definitions considered in
the paper. Section~\ref{sec:methodology} presents the experimental methodology
and used datasets. Results are discussed in Section~\ref{sec:results}, threats
to their validity in Section~\ref{sec:threats}. Before concluding, related work
is discussed in Section~\ref{sec:related}.

\paragraph{Replication package}
A replication package for this paper is available from Zenodo at
\url{https://zenodo.org/record/3610708}.

 \section{What Is a Fork?}
\label{sec:defs}

In this section we explore the spectrum of possible definitions of what
constitutes a \emph{fork}. In the following we will use the term ``fork'' to
mean a forked software \emph{repository}, without discriminating between
``hostile'' (or hard forks, according to the terminology
of~\cite{zhou2019fork}) and development forks. We propose three definitions,
corresponding to three types of forks---type 1 to 3, reminiscent of code clone
classification~\cite{roy2007clonedetectionsurvey,
  rattan2013clonedetectionreview}---along a spectrum of increased sharing of
artifacts commonly found in version control systems (VCS), such as commits and
source code directories.

The first definition, of type 1 forks, relies solely on code hosting platform
information and requires no explicit VCS artifact sharing between repositories
to be considered forks (although it allows it):
\begin{definition}[Type 1 fork, or forge fork]
  \label{def:forge-fork}
  \label{def:type1-fork}
  A repository $B$ hosted on code hosting platform $P$ is a \emph{type 1 fork}
  (or \emph{forge fork}) of repository $A$ hosted on the same platform, written
  $A\forkedI B$, if $B$ has been created with an explicit ``fork repository
  $A$'' action on platform $P$.
\end{definition}

Although informal and seemingly trivial, this definition is both meaningful and
actionable on current major code hosting platforms. For example, GitHub stores
an explicit ``forked from'' relationship and makes it available via its
repositories API:\footnote{\url{https://developer.github.com/v3/repos/},
  retrieved 2020-01-13.}
\begin{quote}
  The \texttt{parent} and \texttt{source} objects are present when the
  repository is a fork. \texttt{parent} is the repository this repository was
  forked from, \texttt{source} is the ultimate source for the network.
\end{quote}
GitLab does the same and exposes type 1 fork information via its projects
API:\footnote{\url{https://docs.gitlab.com/ee/api/projects.html}, retrieved
  2020-01-13}
\begin{quote}
  If the project is a fork, and you provide a valid token to authenticate, the
  \texttt{forked\_from\_project} field will appear in the response.
\end{quote}
which corresponds to exploitable JSON metadata such as:
\begin{lstlisting}[language=json]
{
   "id":3,
   ...
   "forked_from_project":{
      "id":13083,
      "description":"GitLab Community Edition",
      "name":"GitLab Community Edition",
      ...
      "path":"gitlab-foss",
      "path_with_namespace":"gitlab-org/gitlab-foss",
      "created_at":"2013-09-26T06:02:36.000Z",
   ...
\end{lstlisting}

Without getting too formal we observe that each repository is the forge fork of
at most one repository (its parent) and that the relation of being a forge fork
is: not reflexive ($A\notforkedI A$), not symmetric ($A\forkedI B$ does not
imply---and, in fact, excludes---that $B\forkedI A$), not transitive
($A\forkedI B$ and $B\forkedI C$ does not imply---and in fact, due to parent
uniqueness, excludes---that $A\forkedI C$).  The latter might seem surprising
at first but is consistent with the definition, because the action resulting on
the creation of $C$ happened on $B$, not $A$.  (We will introduce later a
related notion of repository relationship that captures transitivity.)

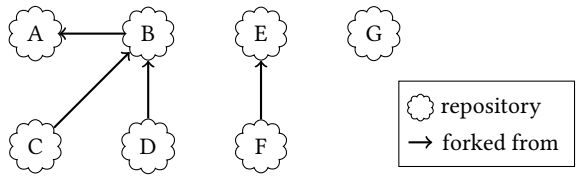
\begin{figure}[t]
  \centering
  \begin{tikzpicture}
	\begin{pgfonlayer}{nodelayer}
		\node [style=origin] (A) at (0, 0)      {A};
		\node [style=origin] (B) at (1.5, 0)    {B};
		\node [style=origin] (C) at (0, -1.5)   {C};
		\node [style=origin] (D) at (1.5, -1.5) {D};
		\node [style=origin] (E) at (3, 0)      {E};
		\node [style=origin] (F) at (3, -1.5)   {F};
		\node [style=origin] (G) at (4.5, 0)    {G};
	\end{pgfonlayer}
	\begin{pgfonlayer}{edgelayer}
		\draw [style=arrow] (B) to (A);
		\draw [style=arrow] (C) to (B);
		\draw [style=arrow] (D) to (B);
		\draw [style=arrow] (F) to (E);
	\end{pgfonlayer}

	\matrix [draw] at (6, -1.2) {
  		\node [style=origin,label=right:repository] {}; \\
  		\node [label=right:forked from,xshift=2pt] {};
        \draw [style=arrow] ++(-0.5em, 0) -- ++(1em, 0) node[right] {}; \\
	};
\end{tikzpicture}
   \caption{Type 1 forks, or forge forks, as declared on code hosting platforms.
    Repository $B$ is a forge fork of $A$, $C$ and $D$ are forge fork of $B$,
    $F$ of $E$, while no repository is a forge fork of $G$. Note how this
    definition induces a global, directed, forge fork graph (specifically: a
    forest of disjoint trees). }\label{fig:forge-fork}
  \label{fig:type1-fork}
\end{figure}

Forge forks induce a global directed graph on repositories, specifically a
forest of disjoint fork-labeled trees, as depicted in \figref{fig:forge-fork}.

Type 2 forks, or \emph{shared commit forks}, are based on the ability offered
by most VCS (and all distributed VCS) of globally identifying commits across
any number of repositories, usually by the means of intrinsic commit
identifiers based on cryptographic hashes~\cite{spinellis2005vcs,
  swhipres2018}. Given the ability to identify commits across different
repositories we can define type 2 forks as follows:
\begin{definition}[Type 2 fork, or shared commit fork]
  \label{def:commit-fork}
  \label{def:type2-fork}
  A repository $B$ is a \emph{type 2 fork} (or \emph{shared commit fork}) of
  repository $A$, written $A\forkedII B$ if it exists a commit $c$ contained in
  the development histories of both $A$ and $B$.
\end{definition}

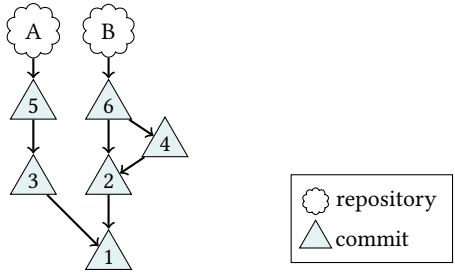
\begin{figure}[t]
  \centering
  \begin{tikzpicture}
	\begin{pgfonlayer}{nodelayer}
		\node [style=origin] (0) at (1, 0) {B};
		\node [style=origin] (1) at (0, 0) {A};
		\node [style=revision] (2) at (1, -1) {6};
		\node [style=revision] (3) at (0, -2) {3};
		\node [style=revision] (5) at (0, -1) {5};
		\node [style=revision] (11) at (1, -2) {2};
		\node [style=revision] (12) at (1, -3) {1};
		\node [style=revision] (13) at (1.75, -1.5) {4};
	\end{pgfonlayer}
	\begin{pgfonlayer}{edgelayer}
		\draw [style=arrow, in=90, out=-90] (0) to (2);
		\draw [style=arrow] (1) to (5);
		\draw [style=arrow] (5) to (3);
		\draw [style=arrow] (11) to (12);
		\draw [style=arrow] (2) to (11);
		\draw [style=arrow] (3) to (12);
		\draw [style=arrow] (2) to (13);
		\draw [style=arrow] (13) to (11);
	\end{pgfonlayer}

	\matrix [draw] at (4.5, -2.5) {
  		\node [style=origin,label=right:repository] {}; \\
  		\node [style=revision,label=right:commit,minimum width=14pt] {}; \\
	};
\end{tikzpicture}
   \caption{Type 2 forks, or shared commit forks. Repository $A$ is a fork of
    $B$ and vice versa, since they share commit $1$.}\label{fig:commit-fork}
  \label{fig:type2-fork}
\end{figure}

\figref{fig:commit-fork} shows an example of 2 repositories, $A$ and $B$ that
are type 2 forks of each other, due to the fact they have in common commit $1$,
the initial commit; their respective development histories diverged immediately
after that commit and never shared any other commits. In the general case
shared commit forks will share many more commits: all the commits that were
available at the time of the most recent development history divergence.

Differently from type 1 forks, the relation of being a type 2 fork is
symmetric ($A\forkedII B$ implies $B\forkedII A$), but still not transitive (as
three repositories $A$, $B$ and $C$ can have shared artifacts between $A$ and
$B$ and between $B$ and $C$ without there necessarily being a shared artifact
between $A$ and $C$).

Intuitively, the notion of shared commit forks is more robust than that of
forge forks because it allows to recognize as forks---in the broad sense of
``repositories that collaborate with one another''---repositories that are
hosted on different platforms. A repository hosted on GitLab.com, or your
personal git repository on your homepage, can be recognized as a fork of a
another hosted on GitHub. The price to pay is that, due to symmetry, the
definition \emph{alone} is not enough to orientate the relation; it does not
capture which repository ``came first''.

We can push this idea further, trying to make it even more robust, and capable
of recognizing as forks repositories that have no recognizable shared commits,
but do share entire source trees. That is of interest when, for example,
collaboration happens using different version control systems (e.g., a
developer using git-svn to participate into the development of a Subversion
based project). Type 3 forks, or \emph{shared root (directory) forks}, allow to
capture those situations:

\begin{definition}[Type 3 fork, or shared root fork]
  \label{def:rootdir-fork}
  \label{def:type3-fork}
  A repository $B$ is a \emph{type 3 fork} (or \emph{shared root fork}) of a
  repository $A$, written $A\forkedIII B$, if there exist a commit $c_A$ in the
  development history of $A$ and a commit $c_B$ in that of $B$ such that the
  full source code trees of the two commits are identical.
\end{definition}

\begin{figure}[t]
  \centering
  \begin{tikzpicture}
	\begin{pgfonlayer}{nodelayer}
		\node [style=origin] (0) at (2, 0) {B};
		\node [style=origin] (1) at (0, 0) {A};
		\node [style=revision] (2) at (2, -1) {10};
		\node [style=revision] (3) at (0, -2) {7};
		\node [style=revision] (4) at (0, -3) {5};
		\node [style=revision] (5) at (0, -1) {9};
		\node [style=directory] (6) at (1, -3) {1};
		\node [style=directory] (7) at (1, -2) {2};
		\node [style=directory] (9) at (3, -1) {4};
		\node [style=directory] (10) at (3, -2) {3};
		\node [style=revision] (11) at (2, -2) {8};
		\node [style=revision] (12) at (2, -3) {6};
	\end{pgfonlayer}
	\begin{pgfonlayer}{edgelayer}
		\draw [style=arrow, in=90, out=-90] (0) to (2);
		\draw [style=arrow] (1) to (5);
		\draw [style=arrow] (5) to (3);
		\draw [style=arrow] (3) to (4);
		\draw [style=arrow] (4) to (6);
		\draw [style=arrow] (3) to (7);
		\draw [style=arrow] (12) to (6);
		\draw [style=arrow] (11) to (12);
		\draw [style=arrow] (2) to (11);
		\draw [style=arrow] (11) to (10);
		\draw [style=arrow] (2) to (9);
		\draw [style=arrow] (5) to (7);
	\end{pgfonlayer}

	\matrix [draw] at (5.5, -2.5) {
  		\node [style=origin,label=right:repository] {}; \\
  		\node [style=revision,label=right:commit,minimum width=14pt] {}; \\
  		\node [style=directory,label=right:root directory] {}; \\
	};
\end{tikzpicture}
   \caption{Type 3 fork, or shared root fork. Repository $A$ is a fork of $B$
    and vice versa, since they share root directory $1$.  As per shared commit
    forks, type shared root forks are symmetric.}\label{fig:rootdir-fork}
  \label{fig:type3-fork}
\end{figure}
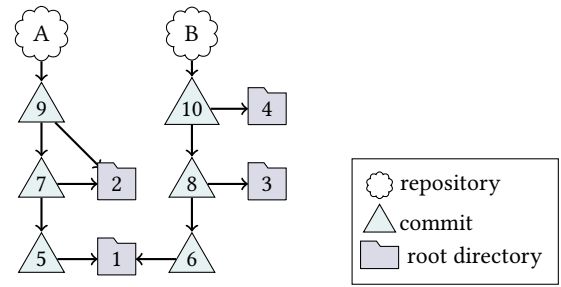

The intuition behind type 3 forks is depicted in \figref{fig:type3-fork}. Note
that it is not enough for the two repositories to share any arbitrary
\emph{sub}-directory to be considered forks, as that would consider as forks
repositories that embed third-party libraries, an arguably undesired
consequence; we need the \emph{root} directories of two commits to be
(recursively) equal for establishing a shared root fork relationship.

The same properties of type 2 forks apply to type 3 forks: the shared root fork
relation is also symmetric. In most VCS, and in all modern DVCS, type 2 forks
is also a strictly larger relation than type 3 forks: $A\forkedII B$ implies
$A\forkedIII B$, because if there exists a shared commit $c$ that makes $A$ and
$B$ shared commit forks, then the root directory pointed by $c$ also makes $A$
and $B$ shared root forks (due to the cryptographic properties of intrinsic
commit identifiers in DVCS).
This property of inclusion, in the sense of one definition implying the other,
is at the heart of the analysis made in section \ref{sec:aggregation-process},
studying the aggregation processes of networks and cliques.

In theory we could go further, and introduce an even more lax notion of fork,
that equates repositories sharing as little as a single file, but that would
exacerbate the problematic behavior we discussed for sharing sub-directories.

Armed with these definitions we will be able to answer RQ1, by comparing the
number of forks identified by Definition~\ref{def:type1-fork} with those
identified by Definition~\ref{def:type2-fork} and/or~\ref{def:type3-fork}
(that we refer to as {\it intrinsic} forks).
To fully address RQ2 on the other hand we need to capture the notion of
``community'' of repositories used for collaboration, as follows:
\begin{definition}[Type $T$ fork network]
  \label{def:fork-network}
  The \emph{type $T$ fork network} of a repository $A$ is the smallest set
  $\network^T_A$ such that:
  \begin{itemize}
  \item $A\in\network^T_A$
  \item $\forall B\in\network^T_A,~B\forked_T C \implies C\in\network^T_A$
  \item $\forall B\in\network^T_A,~C\forked_T B \implies C\in\network^T_A$
  \end{itemize}
\end{definition}

That is, a fork network is the set of all repositories reachable from a given
one, following both forked from (parents) and forked to repositories
(children). The definition is parametric in the type of forks, so we have type
1 fork networks (\networkI), type 2 fork networks (\networkII), and type 3 fork
networks (\networkIII).

A stricter notion that will come in handy is that of repository cliques, sets
of repositories that are all direct forks (i.e., neither transitive nor reverse
transitive) of each other:
\begin{definition}[Type $T$ fork clique]
  \label{def:fork-clique}
  The \emph{type $T$ fork clique} of a repository $A$ is the largest set
  $\clique^T_A$ such that:
  \begin{itemize}
  \item $A\in\clique^T_A$
  \item $\forall C, (\forall B \in \clique^T_A, B\forked_T C \land C\forked_T B) \implies C\in\clique^T_A$
  \end{itemize}
\end{definition}

Note that, while this definition is parametric in the type of forks too, fork
cliques make intuitive sense only for type 2 and type 3 forks; type 1 forks
(forge forks) only have singleton cliques as the relation is not symmetric.

 \section{Methodology}
\label{sec:methodology}

\subsection{Dataset}
\label{sec:dataset}

Our goal is to experimentally determine the amount and structure of forks for
the various definitions we have introduced. To do so we will use two datasets:
the \SWHGD~\cite{swh-msr2019-dataset}, which contains the development history
needed to find intrinsic fork relationships, and a reference forge-specific
dataset, GHTorrent~\cite{GHTorrent}, which contains the fork ancestry
relationships as captured by GitHub.

\paragraph{GHTorrent}
GitHub is the largest public software forge, and is therefore the candidate of
choice to study forge forks (type 1). GHTorrent~\cite{GHTorrent} crawls and
archives GitHub via its REST API and makes periodical data dumps available in a
relational table format. In its database schema, the \texttt{project} table
contains a unique identifier for each repository, and a \texttt{forked\_from}
column contains the ID of the repository it has been forked from if the
repository is considered to be a forge forks.  A single SQL query on this table
allows to extract the full graph of GitHub-declared forks,
e.g.:\footnote{Additional URL gymnastic is needed in the query to
cross-reference GHTorrent project URLs with \SWH ones; we refer to the
replication package for this kind of details.}
\begin{lstlisting}[style=sql]
select parents.url as parent,
       projects.url as child
from projects
inner join projects as parents
      on projects.forked_from = parents.id
\end{lstlisting}

\paragraph{Software Heritage Graph Dataset}
\SWH~\cite{swhipres2017, swhcacm2018} is the largest publicly accessible
archive of software source code and accompanying development history, spanning
more than 90 million software projects retrieved from major development forges
including GitHub and GitLab.com.  The \SWHGD~\cite{swh-msr2019-dataset} is an
offline dataset containing the development history of all the projects in \SWH
in a tabular representation of a unified directed acyclic graph (DAG). As the
archive encompasses a substantial portion of all the public GitHub
repositories, it is possible to cross-reference the origins contained in this
dataset with the ones in GitHub, our reference for forge forks.

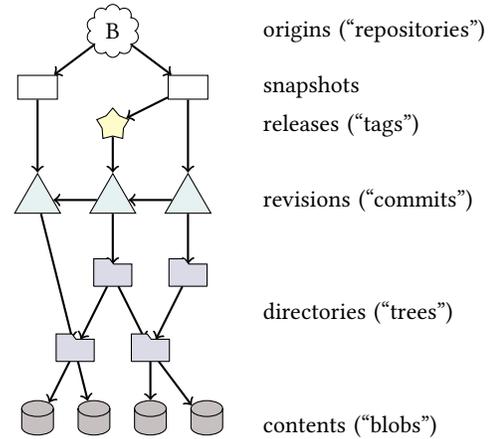
\begin{figure}[t]
  \centering
  \begin{tikzpicture}
	\begin{pgfonlayer}{nodelayer}
		\node [style=origin] (0) at (1, 1.25) {B};
		\node [style=revision] (2) at (2, -1) {};
		\node [style=directory] (9) at (2, -2) {};
		\node [style=directory] (10) at (1, -2) {};
		\node [style=revision] (11) at (1, -1) {};
		\node [style=revision] (12) at (0, -1) {};
		\node [style=directory] (14) at (1.5, -3) {};
		\node [style=directory] (15) at (0.5, -3) {};
		\node [style=content] (16) at (1.5, -4) {};
		\node [style=content] (17) at (2.25, -4) {};
		\node [style=content] (18) at (0.75, -4) {};
		\node [style=content] (19) at (0, -4) {};
		\node [style=release] (20) at (1, 0) {};
		\node [style=snapshot] (21) at (2, 0.5) {};
		\node [style=snapshot] (22) at (0, 0.5) {};
		\node [text width=3cm] (23) at (4.5, 1.25) {origins (``repositories'')};
		\node [text width=3cm] (24) at (4.5, 0.5) {snapshots};
		\node [text width=3cm] (25) at (4.5, 0) {releases (``tags'')};
		\node [text width=3cm] (26) at (4.5, -1) {revisions (``commits'')};
		\node [text width=3cm] (27) at (4.5, -2.5) {directories (``trees'')};
		\node [text width=3cm] (28) at (4.5, -4) {contents (``blobs'')};
	\end{pgfonlayer}
	\begin{pgfonlayer}{edgelayer}
		\draw [style=arrow] (11) to (12);
		\draw [style=arrow] (2) to (11);
		\draw [style=arrow] (11) to (10);
		\draw [style=arrow] (2) to (9);
		\draw [style=arrow] (12) to (15);
		\draw [style=arrow] (10) to (15);
		\draw [style=arrow] (10) to (14);
		\draw [style=arrow] (9) to (14);
		\draw [style=arrow] (15) to (19);
		\draw [style=arrow] (15) to (18);
		\draw [style=arrow] (14) to (16);
		\draw [style=arrow] (14) to (17);
		\draw [style=arrow] (20) to (11);
		\draw [style=arrow] (21) to (2);
		\draw [style=arrow] (21) to (20);
		\draw [style=arrow] (0) to (21);
		\draw [style=arrow] (0) to (22);
		\draw [style=arrow] (22) to (12);
	\end{pgfonlayer}
\end{tikzpicture}
   \caption{The Software Heritage Graph Dataset data model}\label{fig:swh-model}
\end{figure}

The \SWHGD data model maps the traditional concepts of VCS as nodes in a Merkle
DAG~\cite{Merkle}, as shown in \figref{fig:swh-model}. As a consequence, all
the development artifacts, including commits and source trees, are natively
deduplicated within and across projects. This property is particularly useful
to find intrinsic forks, as it enables tracking the relevant artifacts
(revisions and directories) across the entire dataset and link them back to
their source repositories.

The dataset contains two intermediate layers between repositories and the
commit graph they point to: \emph{snapshots}, which are point in time captures
of the state of a repository; and \emph{releases} (or ``tags''), which are
\emph{revisions} (or commits) labeled with a specific name. As none of our fork
definitions depend on these artifacts, the two layers can be flattened out so
that the origins point directly to the revision graph.  Likewise, the blob
layer and the directory layer are not needed to find shared commit forks
(Definition~\ref{def:type2-fork}), while shared root forks
(Definition~\ref{def:type3-fork}) only require the root directory of each
revision.  Filtering out the unnecessary nodes reduces the graph to a more
reasonable size of 2 billion nodes (down from 10 billion), which makes it
easier to process on a single machine. The structure of the resulting subgraph
closely matches the examples in \figref{fig:type2-fork} and
\figref{fig:type3-fork}, making it easy to verify the intrinsic definitions.

We run the experiments on the compressed version of the two graph datasets,
using the WebGraph framework~\cite{boldi-vigna-webgraph-1,
boldi-vigna-webgraph-2}. The \SWHGD is already distributed as a compressed
BVGraph, along with the \texttt{swh-graph} helper
library~\cite{saner-2020-swh-graph} to run traversal algorithms easily.
The GHTorrent can be compressed from its relational database format using the
\texttt{swh-graph compress} utility.

\subsection{Fork networks}
\label{sec:methodology-fork-networks}

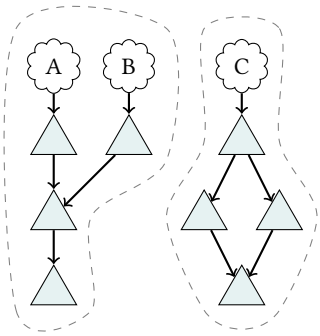
\begin{figure}[t]
  \centering
  \begin{tikzpicture}
	\begin{pgfonlayer}{nodelayer}
		\node [style=origin] (0) at (1, 0) {B};
		\node [style=origin] (1) at (0, 0) {A};
		\node [style=revision] (2) at (1, -1) {};
		\node [style=revision] (3) at (0, -2) {};
		\node [style=revision] (4) at (0, -3) {};
		\node [style=revision] (5) at (0, -1) {};
		\node [style=origin] (6) at (2.5, 0) {};
		\node [style=origin] (7) at (2.5, 0) {C};
		\node [style=revision] (8) at (2.5, -1) {};
		\node [style=revision] (9) at (2, -2) {};
		\node [style=revision] (10) at (3, -2) {};
		\node [style=revision] (11) at (2.5, -3) {};
		\node [style=none] (12) at (-0.5, 0.5) {};
		\node [style=none] (13) at (-0.5, -3.25) {};
		\node [style=none] (14) at (0.5, -3.25) {};
		\node [style=none] (15) at (0.5, -2) {};
		\node [style=none] (16) at (1.5, -1.25) {};
		\node [style=none] (17) at (1.5, 0.5) {};
		\node [style=none] (18) at (2, 0.5) {};
		\node [style=none] (19) at (3, 0.5) {};
		\node [style=none] (20) at (3, -1) {};
		\node [style=none] (21) at (3.5, -2) {};
		\node [style=none] (22) at (3, -3.25) {};
		\node [style=none] (23) at (2.5, -3.5) {};
		\node [style=none] (24) at (2, -3.25) {};
		\node [style=none] (25) at (1.5, -2) {};
		\node [style=none] (26) at (2, -1) {};
	\end{pgfonlayer}
	\begin{pgfonlayer}{edgelayer}
		\draw [style=arrow, in=90, out=-90] (0) to (2);
		\draw [style=arrow] (1) to (5);
		\draw [style=arrow] (2) to (3);
		\draw [style=arrow] (5) to (3);
		\draw [style=arrow] (3) to (4);
		\draw [style=arrow] (7) to (8);
		\draw [style=arrow] (8) to (9);
		\draw [style=arrow] (8) to (10);
		\draw [style=arrow] (9) to (11);
		\draw [style=arrow] (10) to (11);
	\end{pgfonlayer}

\draw [style=cluster] plot [smooth cycle] coordinates {(12.center) (13.center) (14.center)
  (15.center) (16.center) (17.center)};

\draw [style=cluster] plot [smooth cycle] coordinates {(18.center) (19.center) (20.center)
  (21.center) (22.center) (23.center) (24.center) (25.center) (26.center)};
\end{tikzpicture}
   \caption{Fork networks identified as connected components, for the case of
    shared commit (type 2) forks. Connected components are computed on the
    undirected version of the shown Merkle DAG. Measuring network sizes as the
    number of contained repository nodes we obtain that: repositories A and B
    are forks of each other and members of a network of size 2, while
    repository C is in its own singleton network.}
  \label{fig:fork-clusters}
\end{figure}

The easiest way to get a first sense of the amount and structure of forks
according to the various definitions is to find all fork networks, as per
Definition~\ref{def:fork-network}. This can be done in linear time with a
simple graph traversal with linear complexity: two repositories are in the same
network if and only if there exists a path between them in the undirected
subgraph of origins and revisions. (We recall from the dataset section that we
have removed the snapshot and revision layers, so that root commits are
directly pointed by repository nodes.) Finding all the fork networks is
therefore equivalent to computing the connected components on this subgraph, as
exemplified in Figure~\ref{fig:fork-clusters}.

Using fork networks has the advantage of allowing easy interpretations of the
results. First, it is trivial to quantify how many repositories are forks by
counting the number of repositories that belong to non-singleton networks.
Besides, a direct comparison can be made between the distribution of forge
forks and shared commit or root forks, as networks provide a partition method
for both graphs. The sizes of the networks can be directly compared between
the three definitions while keeping the invariant of number of total
repositories.
This is not the case when looking at fork cliques, since the same repository
can be found in multiple cliques, which makes comparison harder.

In GHTorrent origins are already linked together in a global graph where the
edges represent the forge-level forking relationships. We can partition this
forge fork graph in fork networks similarly by computing all its connected
components.

Our experimental design is therefore as follows: first, we list the common
non-empty repositories between the \SWHGD and GHTorrent. We then extract the
aforementioned subgraphs: the development history graph for \SWH (origins
$\to$~\{revisions, releases\} $\to$~commits) and the fork graph for GHTorrent
(origins $\to$~origins). We then compute the connected components of each graph
using a simple depth-first traversal algorithm, then output the origins
contained in each component.

\subsection{Fork cliques}\label{sec:methodology-fork-cliques}

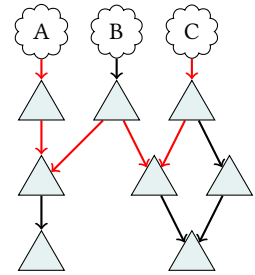
\begin{figure}[t]
  \centering
  \begin{tikzpicture}
	\begin{pgfonlayer}{nodelayer}
		\node [style=origin] (0) at (1, 0) {B};
		\node [style=origin] (1) at (0, 0) {A};
		\node [style=revision] (2) at (1, -1) {};
		\node [style=revision] (3) at (0, -2) {};
		\node [style=revision] (4) at (0, -3) {};
		\node [style=revision] (5) at (0, -1) {};
		\node [style=origin] (6) at (2, 0) {};
		\node [style=origin] (7) at (2, 0) {C};
		\node [style=revision] (8) at (2, -1) {};
		\node [style=revision] (9) at (1.5, -2) {};
		\node [style=revision] (10) at (2.5, -2) {};
		\node [style=revision] (11) at (2, -3) {};
	\end{pgfonlayer}
	\begin{pgfonlayer}{edgelayer}
		\draw [style=arrow, in=90, out=-90] (0) to (2);
		\draw [style=red arrow, in=90, out=-90] (1) to (5);
		\draw [style=red arrow] (2) to (3);
		\draw [style=red arrow] (5) to (3);
		\draw [style=arrow] (3) to (4);
		\draw [style=red arrow] (7) to (8);
		\draw [style=red arrow] (8) to (9);
		\draw [style=arrow] (8) to (10);
		\draw [style=arrow] (9) to (11);
		\draw [style=arrow] (10) to (11);
		\draw [style=red arrow] (2) to (9);
	\end{pgfonlayer}
\end{tikzpicture}
   \caption{Example of misleading clustering of fork networks.  Here,
    repositories A and C are in the same network because there is a path
    between them, even though they do not share common development history.}\label{fig:fork-transitive-fail}
\end{figure}

While partitioning the corpus in fork networks gives a good idea of how
intrinsic forks are linked together, it can group together repositories that
are not forks of each other, as the intrinsic fork relationship is not
transitive.  Figure~\ref{fig:fork-transitive-fail} shows a pattern, that we
have verified as commonly found in the wild, where two different cliques will
be merged in the same fork network---A and B are part of the same clique as
they share development history; the same applies to B and C; whereas A and C do
not share any part of their respective development histories but will end up in
the same network. We expect this effect to merge cliques into giant components,
that will make the size of the largest networks hard to interpret.

The other interesting metric that can be looked at is the distribution of fork
cliques, as defined in Definition~\ref{def:fork-clique}. While cliques do not
provide a partition function for the graph, they allow to narrow down the
actual extent of forking relationships within large fork networks.

Due to the fact that shared commit fork cliques are defined pairwise, the naive
algorithm to find all the inclusion-maximal cliques is superlinear: for each
repository, walk through its commit history and add all the commits to a queue,
then take the transposed graph to walk through the commit history backwards and
list all repository leaves. The time complexity of this algorithm is highly
unpractical: in the worst case, if all the repositories are forks of each
other, is has time complexity of $\mathcal{O}(R \times C)$ where $C$ is the
number of commits and $R$ the number of repositories in the graph.

However, clever use of some properties on the DAG structure of the commit graph
can substantially speed up the algorithm. First, fork cliques can be generated
by iterating on the common ancestors instead of the repositories: for each
commit $c$, if it has more than one repository leave when doing a traversal on
the transposed graph, then $c$ was a common commit ancestor, and the generated
set of repositories is a fork clique. Besides, since the ancestry relationship
is transitive, the clique with commit $c$ as a common ancestor is the same as
the clique generated by running the traversal on its parents. By induction, it
is possible to compute all the cliques simply by doing one traversal per
``root'' commit.

\begin{algorithm}[t]
  \caption{Find all the fork cliques}\label{algo:clique}
  \begin{algorithmic}
\Function{FindOriginLeaves}{r}
    \State $S_O \gets$ empty set
    \ForAll{$n \in \Call{AncestorsDFS}{r}$}
        \If {\Call{type}{n} $=$ \texttt{ORIGIN}}
            \State add $n$ to $S_O$
        \EndIf
    \EndFor
    \State \Return $S_O$
\EndFunction

\Function{FindCliques}{$G$}
    \State $S_F \gets$ empty set
    \State $S_C \gets$ empty set
    \ForAll{$n \in G$}
        \If {$\Call{type}{n} = \texttt{REVISION}~\textbf{and}~n~\text{has no
        parents}$}
            \State $c \gets \Call{FindOriginLeaves}{n}$
            \State $f_c \gets \Call{Fingerprint}{c}$
            \If {$f_c \centernot\in S_F$}
                \State add $f_c$ to $S_F$
                \State add $c$ to $S_C$
            \EndIf
        \EndIf
    \EndFor
    \State \Return $S_C$
\EndFunction

\end{algorithmic}
 \end{algorithm}

The resulting algorithm is Algorithm~\ref{algo:clique}: for each root commit
with no parents, we generate the clique of all repositories that contain it. We
use a cryptographic hash fingerprint to avoid adding multiple times the same
clique if it has multiple root commits. While this algorithm technically does
not change the worst case complexity on arbitrary graphs, it is still a huge
speed improvement in our case, as commit chains tend to be degenerate (i.e.,
very long chains with indegrees and outdegrees close to 1 on
average). Algorithm~\ref{algo:clique} has a best-case complexity of
$\Theta(C)$, equivalent to a single DFS traversal. The commit graph is largely
close to this best-case scenario, making the algorithm run in just a few hours
on the entire corpus.

While Algorithm~\ref{algo:clique} works well for shared commit forks, the
speedup does not apply to shared root forks: the induction property no longer
works for root directories, as they are not organised in nearly-degenerate
chains. The time complexity for type 3 forks is closer to the worst case of
$\mathcal{O}(C \times R)$, which makes the clique analysis impractical for this
kind of forks.

\paragraph{P-clique partition function}
While cliques do not directly provide a way to partition the corpus in several
fork clusters (because a single origin can be contained in multiple cliques),
it is possible to define a partition function based on them, e.g., by always
assigning repositories to the largest clique they belong to. As repositories
belonging to multiple cliques appear to be a quite rare occurrence (as they
require the equivalent of a \texttt{git merge -{}-allow-unrelated-histories} on
two completely different repositories), the arbitrary criterion choice is not
expected to be a significant caveat to interpret the results.

We use the Algorithm~\ref{algo:clique-partition} to generate the partition
function of the graph based on cliques. To implement the criterion of
attributing repositories to their largest cliques, cliques are processed in
decreasing order of size. Building a reverse index of ``repository
$\rightarrow$ clique it belongs to'' allows direct access to the subsequent
occurrences of repositories in smaller cliques to remove them. After doing so,
the cliques left empty are removed and the newly generated graph partition can
be returned.

The output of this algorithm generate a set of sets of origins that are subsets
of the input fork cliques. We call this set \emph{``fork p-cliques''} to
emphasize the fact that they form a partition of the repository set in which
all the groups are fork cliques.

\begin{algorithm}[t]
    \caption{Compute the p-cliques partition function}
    \label{algo:clique-partition}
    \begin{algorithmic}
    \Function{CliquesToPartition}{$L_C$}
        \State $\Call{ReverseSizeSort}{L_C}$ \Comment{Process larger cliques
        first}
        \State $I \gets$ empty map \Comment{Build reverse index}
        \ForAll{$c_i \in L_C$}
            \State add \{$i \rightarrow c_i$\} to $I$
        \EndFor
        \ForAll{$c_i \in L_C$}
            \ForAll{$r_j \in c_i$}
                \ForAll{$s \in I[r_j]$}
                \Comment{\parbox[t]{.32\linewidth}{
                    Remove subsequent occurrences of $r_j$}}
                    \vspace{-\baselineskip}
                    \If{$k > i$}
                        \State remove $r_j$ from $s$
                    \EndIf
                \EndFor
            \EndFor
        \EndFor
        \State $L_C \gets \Call{RemoveEmptySets}{L_C}$ \Comment{Remove cliques left empty}
        \State \Return $L_C$
    \EndFunction
    \end{algorithmic}
\end{algorithm}

Once this p-clique graph partition is established, the fork definition can once
again be compared with the forge definition, by looking at the difference
between the size distribution of the partitioned cliques of type 2 forks and
the size distribution of networks for type 1 forks.

 \section{Results}
\label{sec:results}

We identified 71.9\,M repositories in common between the \SWHGD and GHTorrent,
41.4\,M of which are non-empty. We focused our experiments on these
repositories.

\subsection{Fork networks}
\label{sec:results-networks}

In the GHTorrent graph, we found 25.3 M different connected components, among
which 22.9 M repositories isolated in their own component, which means they are
not forge forks (type 1) of other repositories.  The other 2.4 M connected
components contain the remaining 18.5 M repositories, which are all in fork
networks.  These forge forks represent 44.74\% of all repositories.

In the \SWHGD, we found 24.0 M connected components, among which 21.3 M
isolated repositories. The remaining 2.6 M components contain 20.1 M shared
commit forks (type 2), i.e., 48.44\% of all repositories. We have hence almost
9\% \emph{more} shared commit forks than forge forks, which is a significant
divergence for the most strict definition of forks based on shared VCS
artifacts.

\begin{table}[t]
    \centering
    \caption{Number of forks and networks by fork type}\label{tab:network-results}
    \begin{tabular}{l|c|c}
      \textbf{Fork type} & \textbf{\# forks} & \textbf{\# networks} \\
      \hline
      Forge forks (type 1)         & 18.5 M (44.7\%) & 25.3 M \\
      Shared commit forks (type 2) & 20.1 M (48.4\%) & 24.0 M \\
      Shared root forks (type 3)   & 25.3 M (61.1\%) & 18.5 M \\
    \end{tabular}
\end{table}

For shared root forks (type 3), we found 18.5 M connected components, among
which 16.1 M isolated repositories and 25.3 intrinsic forks (61.08\% of all
repositories), which is almost 37\% more than the forge forks. These results
are summarized in Table~\ref{tab:network-results}.  They suggest that
\textbf{in between 1.6 M (3.8\% of total) and 6.8 M (16\%) repositories might
  be overlooked when studying forks using only GitHub metadata} as a source of
truth for what is a fork.

\begin{figure}[t]
    \centering
    \includegraphics[width=\linewidth]{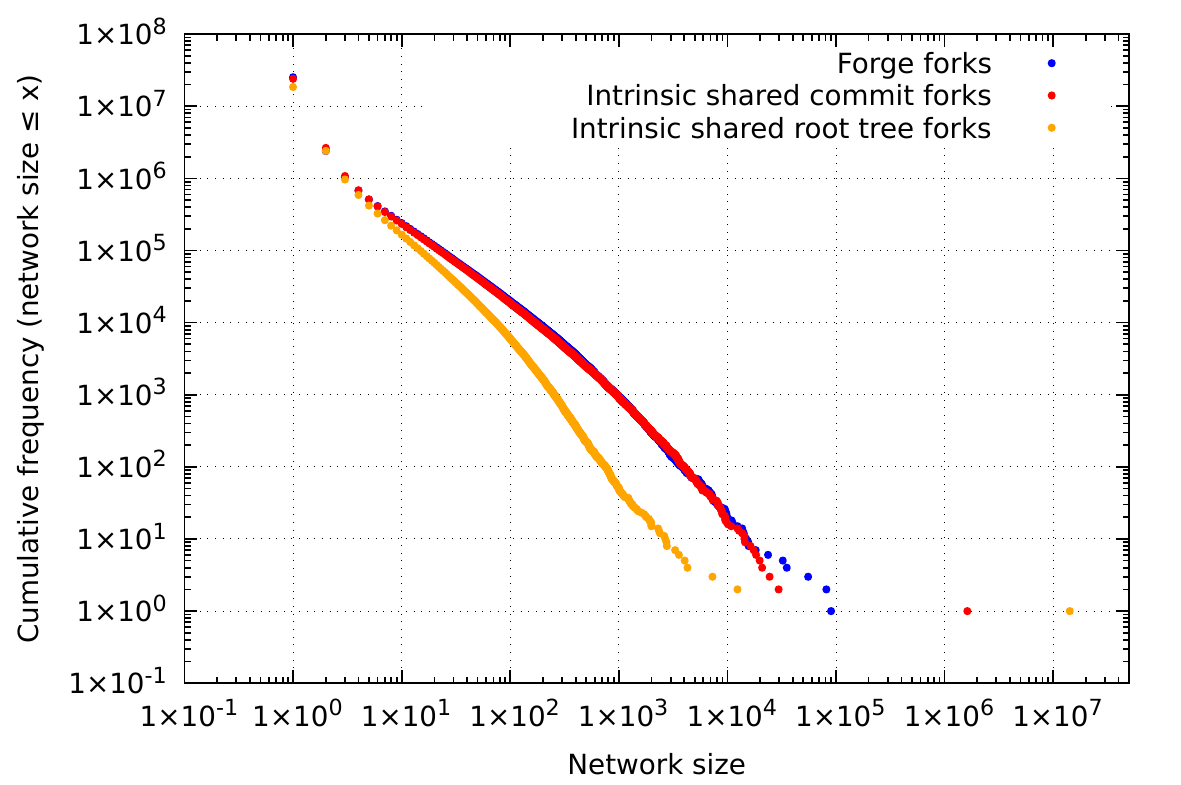}
    \caption{Cumulative frequency distribution of fork network sizes}\label{fig:fork-network-freq-distrib}
\end{figure}

\figref{fig:fork-network-freq-distrib} shows the cumulative frequency
distribution of fork networks for intrinsic forks and forge forks. That is, for
each fork network size $x$, the number of repositories in networks of size
$\geq x$ is shown. At first glance, the distribution of forge forks and shared
commit forks appear to be pretty similar (although the log scale minimizes the
differences between the two), which is a good sign that the two definitions are
not returning vastly different results. The average size of fork networks is
also about the same ($\approx 7.6$ for type 2 forks, $\approx 7.7$ for type 1).
The situation appears to be quite different for shared root forks, where the
average size is $\approx 10.5$ and the frequency distribution is significantly
farther from the reference distribution of forge forks.

One distinguishing feature of each distribution of type 2 and type 3 forks is
the size of the largest connected component, which is significantly larger than
the largest networks of forge forks (by a factor of 17 for shared revision
forks, and 157 for shared root forks). As discussed in
Section~\ref{sec:methodology-fork-cliques}, this is an expected outcome of our
use of network as a quantification metric and confirms the need for further
analysis through fork cliques.  This does not however have any implications on
the quantification aspect of the experiment, as partitioning this network
further using fork cliques would still yield the same number of non-isolated
repositories.

\subsection{Fork cliques}

As expected, running Algorithm~\ref{algo:clique} to generate the
shared-revision cliques on the compressed graph does not take more than an
hour, which is the same order of magnitude as the time needed for a simple full
traversal of the revision graph~\cite{saner-2020-swh-graph}. This confirms our
prediction that in the shared-revision case, the average-case runtime of the
algorithm is close to $\Theta(R)$.

The algorithm finds 24.5 M cliques, although the results are difficult to
interpret in this current state as the cliques overlap together. A few
key observations can nevertheless already be made, notably the absence of very
large cliques: the largest clique contains 92.4 M repositories, which is very
similar to the largest forge fork network (which contains 90.2 M repositories).
This is consistent with our intuition expressed in
Section~\ref{sec:methodology-fork-cliques} that the largest intrinsic fork
networks are a specific feature of networks (as seen in
Figure~\ref{fig:fork-transitive-fail}), and that these artifacts disappear when
looking at the cliques. It is also possible to measure how the cliques overlap:
28 M repositories are present in a single clique, while the remaining 13.3 M
appear two times or more. On average, each repository appears in $\approx 1.47$
cliques.

Computing the p-clique partition function using
Algorithm~\ref{algo:clique-partition} removes this overlap to allow a direct
comparison with the forge fork networks. This algorithm takes a few minutes to
process the 24 million cliques and returns the p-clique partition directly,
restoring the invariant of total number of repositories (41.5 M).

There are 24.0 M of p-cliques partitioning the graph, which is pretty close to
the number of forge fork networks in GitHub (25.3 M).
21.3 M repositories are isolated in their own p-clique (51.6\%), and the
remaining 48.4\% are in cliques of size larger than one, which is consistent
with the findings of Section~\ref{sec:results-networks} which uses fork
networks as a quantification mechanism.

\begin{figure}[t]
    \centering
    \includegraphics[width=\linewidth]{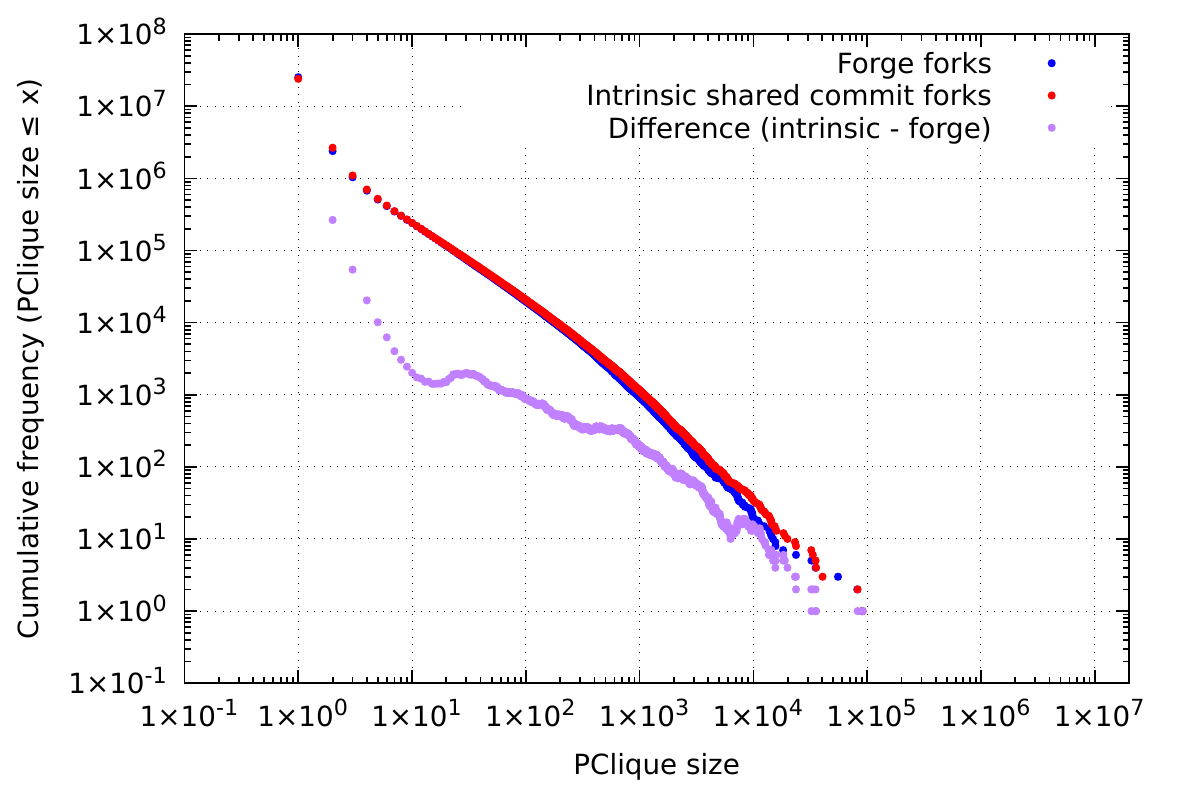}
    \caption{Cumulative frequency distribution of intrinsic fork p-cliques
    compared to forge fork networks}\label{fig:fork-clique-freq-distrib}
\end{figure}

\figref{fig:fork-clique-freq-distrib} shows the cumulative frequency
distribution of the sizes of the shared-commit fork cliques, compared to the
baseline of forge fork networks. As before, the graph can be read as: ``for
each clique (resp.\ forge fork network) of size $x$, the number
of repositories found in cliques (resp.\ networks) of size $\geq x$''.

The visual similarity between the two distributions is striking: while the
shared-commit p-clique distribution seems to be consistently above the
forge-fork network baseline for groups of size $\geq 2$, they always appear to
be very close to each other, even farther in the tail. This \textbf{suggests
that type 2 forks capture well what developers typically recognize as forks}.

To formally assess this similarity, the graph also exhibits the cumulative
difference between the clique distribution and the baseline. This is in
essence, the cumulative size distribution of the cliques of forks
\emph{overlooked} when using only the GitHub metadata. 
This cumulative
distribution \textbf{mostly stays positive, suggesting that using DVCS data to
identify forks is overall a net gain in coverage}.
It also appears that the
difference is typically at least one order of magnitude less than the size of
the clusters, emphasizing the proximity between the two definitions.

\subsection{Aggregation process}
\label{sec:aggregation-process}

Two repositories having a common commit ancestor necessarily have a common root
source tree (the root source tree of that common commit ancestor), so all the
repositories that belong to the same shared-commit network also belong to
the same shared-directory network.
Similarly, we expect that most origins declared as forge forks will be
in the same shared commit and shared root source tree fork networks.
By switching from one definition to another, we expect the clusters to
aggregate together smaller clusters from the previous definitions.

To characterize this aggregation process into fork clusters at different
granularities, we compute the Kolmogorov-Smirnov (KS) distance between the
weighted cumulative distributions function of the clique or network size.

We note $\delta O$ the KS difference between a fork
definition A and a fork definition B, and
represent it as a function of the size of the network (or partitioned clique).
By definition $\delta O$ is always equal to zero for sizes $s~=~1$ (as all the
forks are in clusters of size $s \geq 1$) and $s~=~\max(\text{cluster
sizes})$ (as there are no clusters larger than this size).

\begin{figure}[t]
    \centering
    \includegraphics[width=\linewidth]{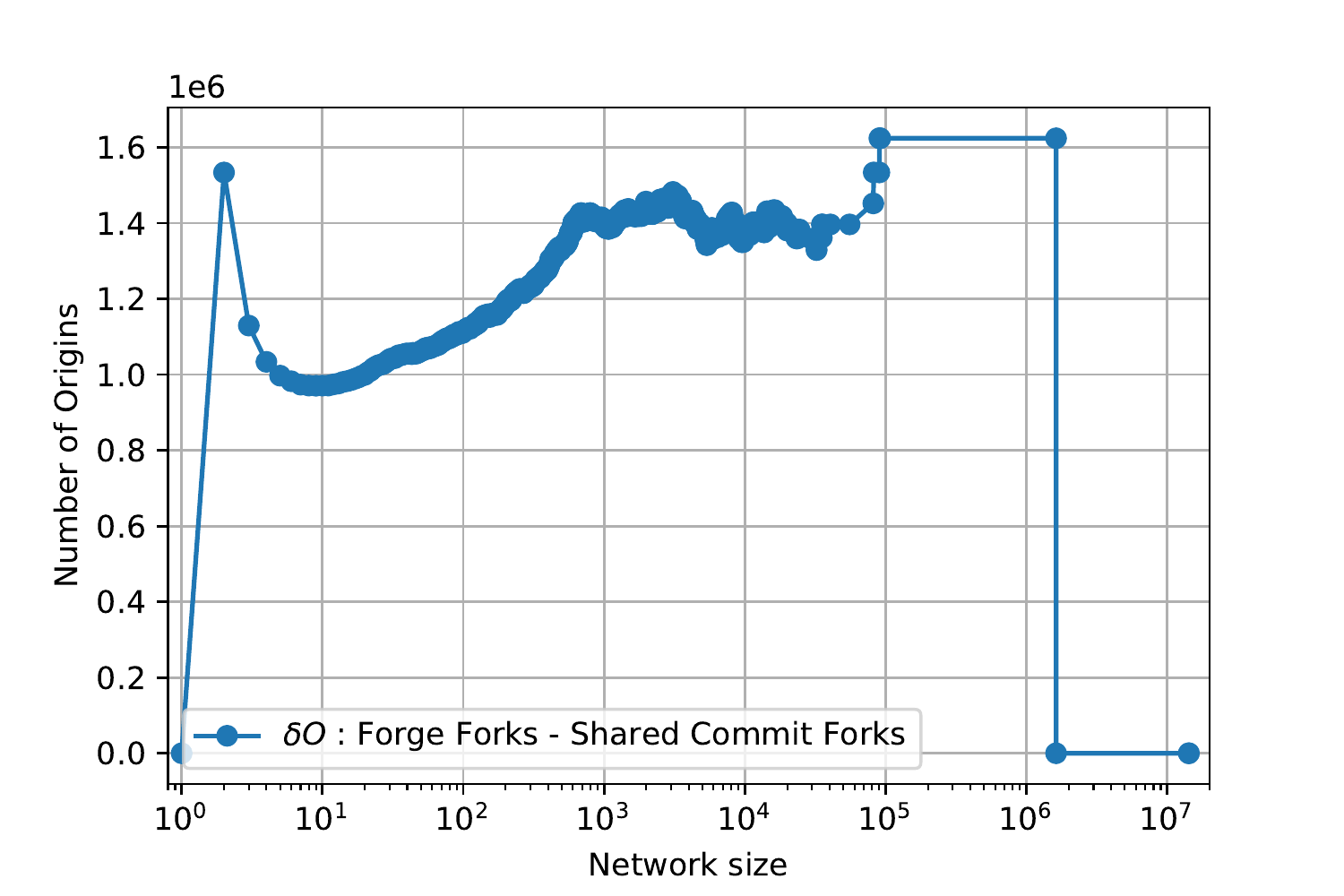}
\includegraphics[width=\linewidth]{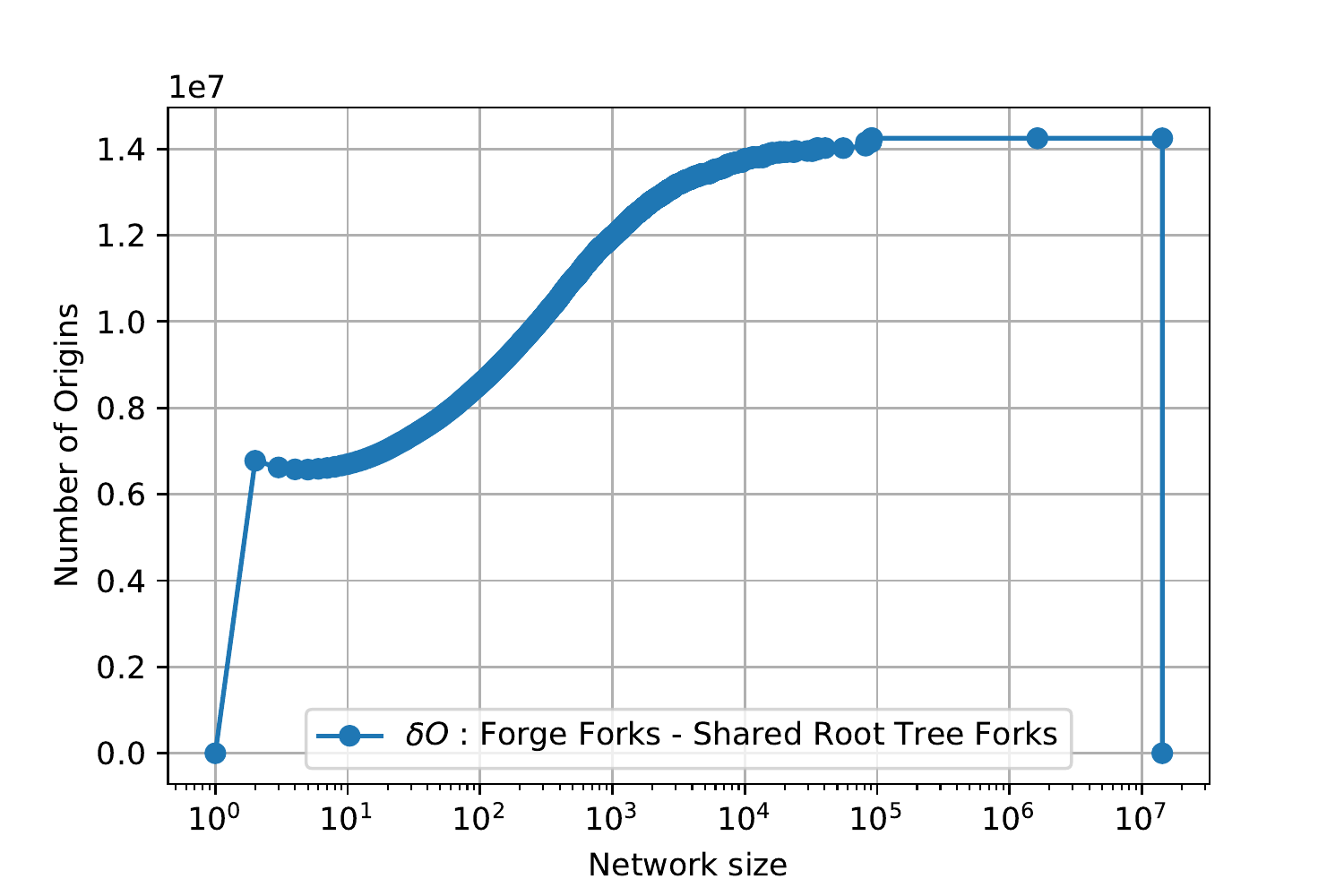}
\includegraphics[width=\linewidth]{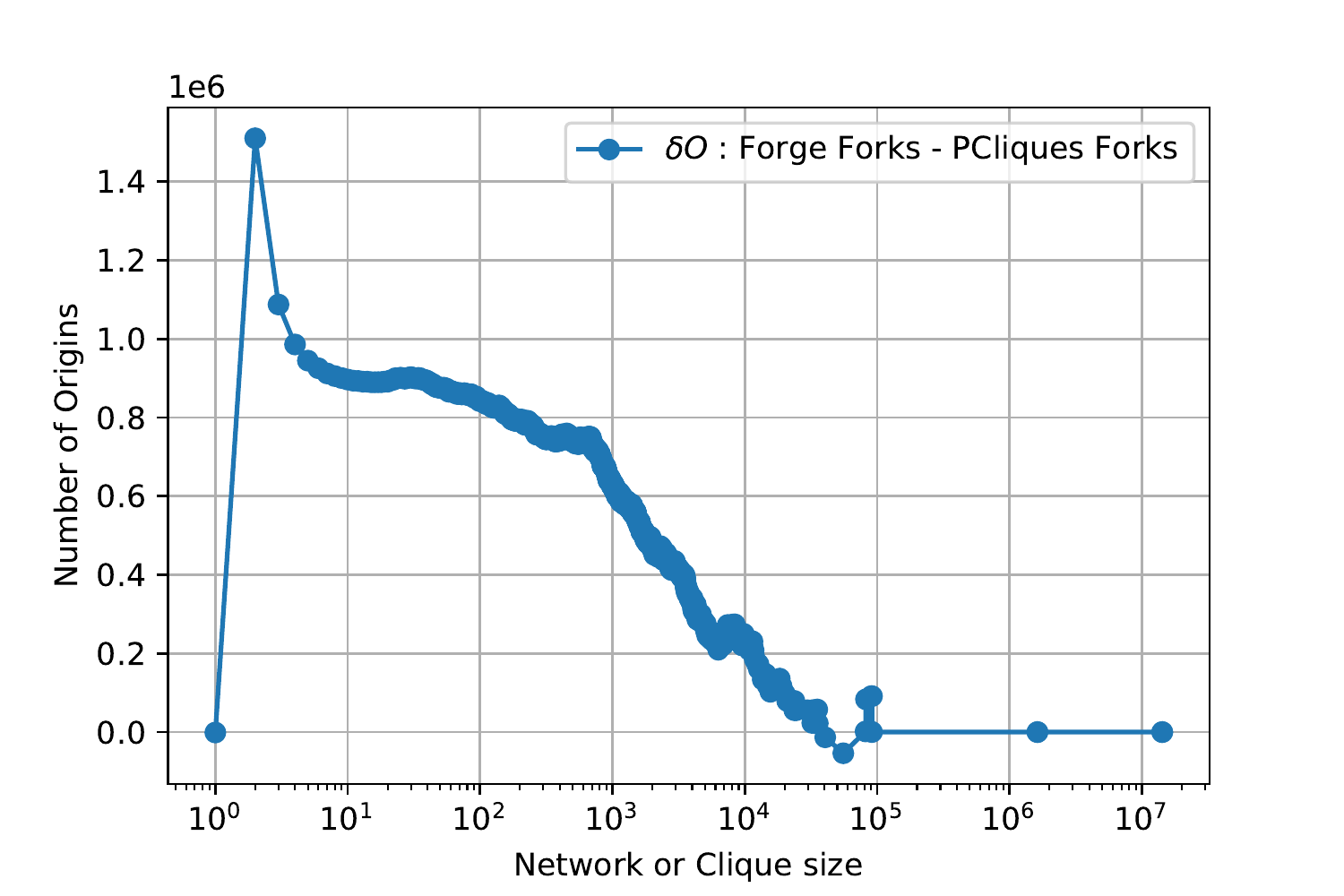}
\caption{Complementary Cumulative Weighted Distribution Functions
    Differences between forge fork network and shared-commit fork network
    (top), shared root source tree fork network (middle), and p-cliques
    based fork network (bottom).}
    \label{fig:flux-gh-rev_rootdir_pcliques}
\end{figure}

Because the total number of repositories is invariant, we can plot the KS
distance weighted by repositories to see how the repositories found in fork
networks (or cliques) of a given size will progressively aggregate into fork
networks (or cliques) of different sizes.
\figref{fig:flux-gh-rev_rootdir_pcliques} represents $\delta O$ between
the forge fork definition baseline and: shared commit fork networks (top),
shared commit p-cliques (bottom), and shared root tree fork networks (middle)).

While this analysis shows the flux of repositories between clusters identified
by the different definitions, it can mask some compensating phenomena by
merging independent processes, as some repositories can migrate from larger to
smaller clusters, 
sometimes leading to $\delta O<0$ (\figref{fig:flux-gh-rev_rootdir_pcliques}, bottom, $size\sim10^5$).

\begin{figure}[t]
    \centering
\includegraphics[width=\linewidth]{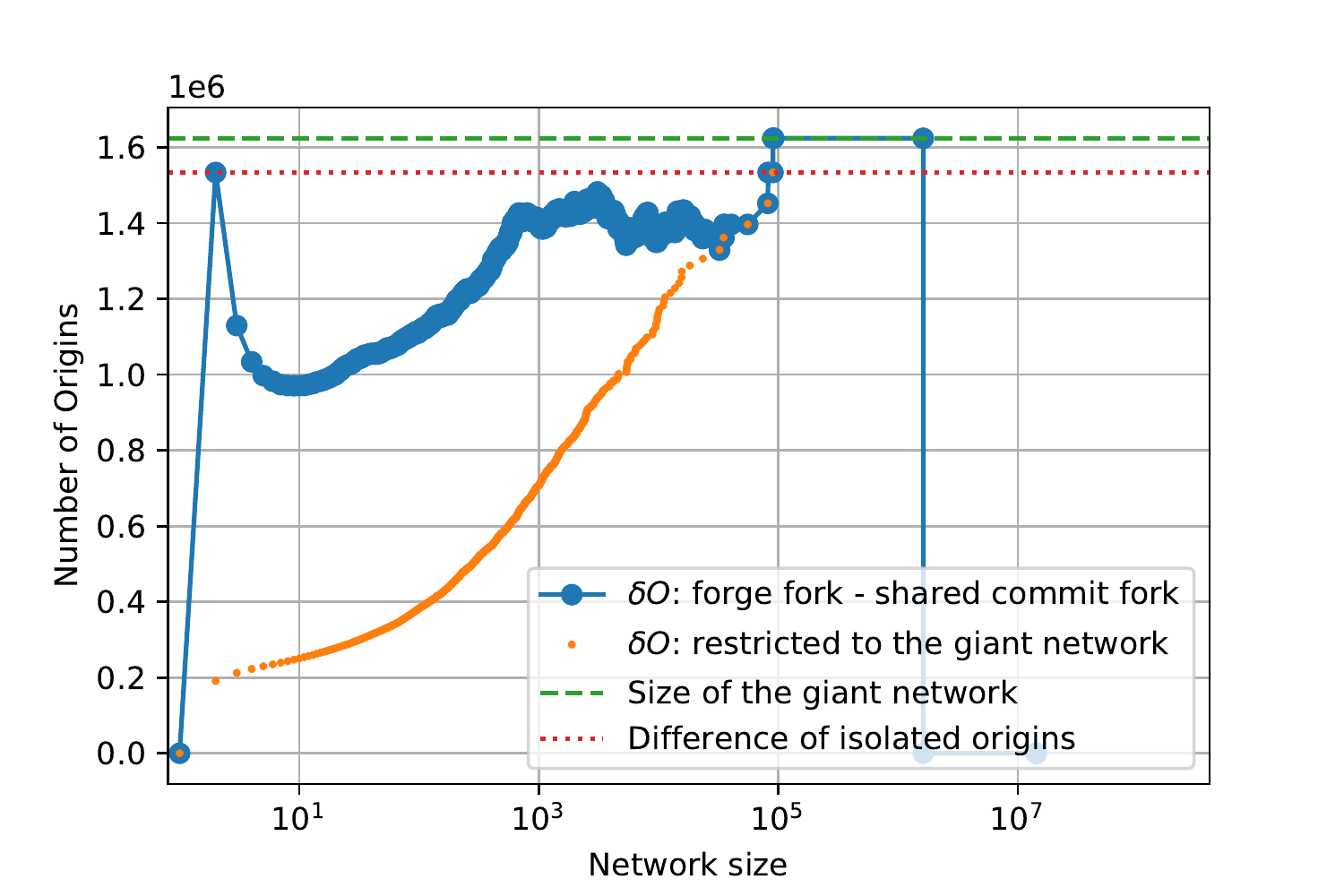}
\caption{Contribution (orange dots) of the giant (largest) network that
    appear using shared commit fork definition w.r.t.
    $\bm{\delta O :} \textbf{forge~fork} - \textbf{shared~commit~fork}$ (Same as
    Fig.~\ref{fig:flux-gh-rev_rootdir_pcliques}, Top)}
    \label{fig:Diff_WCCDF_all}
\end{figure}

To narrow down this phenomenon, we specifically focus on the
largest shared-commit fork network to see how it contributes to the global
flux. By taking the repositories in this network and the size
distribution of the forge fork networks, we show in \figref{fig:Diff_WCCDF_all}
the repository flux, as defined above, and compare it to the corresponding
global flux.

Several points are noteworthy.
First, only $15\%$ (200 k origins over 1.53 M (red dotted)) of the
origins that were isolated (blue dot for $size=2$) in forge fork network
are aggregated in the giant network of shared commit fork.
This shows that the aggregation mechanism is not only to this "giant" network,
since $\sim 85\%$ of the origins are aggregated within smaller networks.

Then, the flux for the 6 largest networks (isolated blue and orange dots around
$10^5$ that overlap) is almost the same whether we restrict ourselves to the
origins of the giant network (orange line) or to all the networks (blue line).
We conclude that aggregation for the large network sizes is dominated by
absorption into the giant cluster, without any redistribution to smaller
networks.

This confirms that the ``aggregation/merge'' mechanism which happens when
changing the definition is not just an absorption phenomenon into a ``super
attractor'', but concerns all network sizes, with larger networks absorbing
networks of any size.

 \section{Threats to validity}
\label{sec:threats}

\paragraph{Internal validity}
Aside from forge forks (type 1) we have no certainty on how well the proposed
fork definitions capture what developers would recognize as forks. While shared
commit (type 2) and shared root (type 3) forks make intuitive sense, the sheer
volume of data to be analyzed makes it very hard to rule out the existence of
pathological cases. Certainly type 2 and type 3 fork definitions can be
``gamed'', making unrelated repositories appear as forks when they intuitively
are not. Unusual development workflows might also induce topology in the global
development graph that merge together repositories that would not be considered
forks by developers. There exists an apparent trade-off here between fully
automatable definitions based on VCS artifact sharing, and qualitative
assessment by developers that does not scale to datasets like the one studied
here.

Also as a consequence of the above we do not feel confident at this stage in
making a methodological judgment call on whether type 2 or type 3 fork
definitions ``better'' capture the essence of a fork. We simply warn scholars
about the extent of the discrepancies between the number of forks detectable
via shared VCS artifacts and forge-level metadata. Further work, of both
statistical nature (looking for outliers) and based on structured interviews
with developers (to review uncommon cases), is needed to improve over this
point.

\paragraph{External validity}
The datasets used in this study do not capture the full extent of publicly
available development history, notably due to their snapshot nature and their
assembly through periodic crawling processes. The \SWHGD only contains data
from various forges to the extent of what is covered by the \SWH archive, which
might lag behind the tracked forges, and GHTorrent is GitHub-specific. As we
need comparable samples we were limited by the intersection of the two datasets
in this study, which composes these limitations. Still, to the best of our
knowledge this is one of the largest quantitative fork study to date, having
considered more than 40 M public version control system repositories.

In the future it would be interesting to extend this approach to forges that
are raising in popularity, and most notably GitLab. For that we would need a
GHTorrent equivalent (or corresponding ad hoc crawling of that forge for the
purposes of the study only).

 \section{Related work}
\label{sec:related}

Accounts of the history of forking have been given by
Nyman~\cite{nyman2016forkhistory} and Zhou et al.~\cite[Section
2]{zhou2019fork}. The latter also covers the terminological and cultural shift
from hard forks (to be avoided) to forking as the mere technical act of
duplicating VCS history, possibly as basis for future collaboration. The
present work is agnostic to which interpretation prevails, as in both cases the
main observable effect of forking are VCS repositories that share parts of an
initially common development history.

\paragraph{Hard forks}

Hard forks have been studied extensively in seminal work by
Nyman~\cite{nyman2011-fork-or-not, nyman2012forking-sustainability,
  nyman2014forking-hackers, nyman2016forkhistory}, covering historical origins,
motivations for forking (or not), and sustainability considerations in the
socioeconomic context of free/open source software (FOSS) development.  Robles
and Barahona~\cite{robles2012forks} give a detailed account of famous hard
forks, covering history, reasons, and outcomes.

These and other studies of hard forks are qualitative and focused in nature.
  This paper is complementary to
them as it proposes tools to identify and quantitatively measure and observe
forks, addressing the need of more extensive and homogeneous fork research
already observed in~\cite{robles2012forks}. As far as we could determine
without fully replicating the corresponding studies, the VCS repositories
involved in the hard fork cases cited thus far would be correctly identified as
either type 2 or type 3 forks.

\paragraph{Development forks}

With the advent of DVCS and social coding~\cite{lima2014ghsocial}, a
significant amount of empirical research has been devoted to development
forks. Motivations for forking on GitHub have been studied by Jiang et
al.~\cite{jiang2017whyfork}.

The structure of forks on GitHub has been analyzed in several studies. Thung et
al.~\cite{thung2013network} have characterized the network structure of social
coding of GitHub, including forks. Padhye~\cite{padhye2014extcontrib} have
measured external contributions from non-core developers. Biazzini and Baudry
have proposed metrics to quantify and classify collaboration in GitHub
repositories pertaining to the same fork tree~\cite{biazzini2014maythefork}.
Rastogi and Nagappan~\cite{rastogi2016forking}---as well as Stanciulescu et
al.~\cite{stuanciulescu2015forked} for firmware projects---have characterized
forks on GitHub based on the flow of commits between them and the originating
repository.

Various performance aspects of the pull request development
model~\cite{gousios2014pullrequests, gousios2015pullrequests} have been also
studied. Latency in acceptance has been a popular one~\cite{yu2015waitforit,
  tsay2014influence}; the amount of generated community
engagement~\cite{dabbish2012socialcoding, dabbish2012transparency} another one.
A more general accounting of efficient forking patterns has recently been given
by Zhou et al.~\cite{zhou2019fork}.

To the extent we could determine it without full replication, all
aforementioned studies on forking for social coding purposes rely on platform
(and more specifically GitHub's) metadata to determine which repository is a
fork (of which other). As such, involved repositories would be recognized as
type 1 forks, and non type 1 forks (but nonetheless type 2 or 3 forks) might
have been overlooked in the studies. To be clear: we have no reason to believe
that the findings in those studies would turn out to be different by enlarging
the set of considered forks using the alternative definitions. We simply
propose to acknowledge fork type discrepancy as an internal validity threat in
future studies.

\paragraph{Fork definitions}

Aside from the already discussed hard forks v.~development forks distinction,
the only other work we are aware of on formal or semi-formal fork
characterization is~\cite{swh-provenance-tr}, which introduces the notion of
\emph{most fit fork}: a repository that, within a group of VCS repositories
that share commits, contain the largest number of commits. The notion is
proposed as a long-term approximation of the main development line of a forked
(hardly or otherwise) project. Our notion of type 2 fork clique captures the
same idea; additionally we show how to use it to partition the global set of
VCS repositories into independent clusters instead of partitioning the global
set of commits.

\paragraph{Methodology}

Methodological issues and risks in analyzing GitHub were pointed out by
Kalliamvakou et al.~\cite{kalliamvakou2014promises}. While not directly
addressed as an explicit risk, forks not recognized as such are echoed by
perils \emph{``I: A repository is not necessarily a project''} and \emph{``IX:
  Many active projects do not conduct all their software development in
  GitHub''} in that article. Proposed mitigations were, respectively,
\emph{``consider the activity in both the base repository and all associated
  forked repositories''} and \emph{``Avoid projects that have a high number of
  committers who are not registered GitHub users and projects which explicitly
  state that they are mirrors in their description''}.

Half a decade later it is arguably \emph{less} of a risk that development
happens elsewhere and that a high number of committers are not registered
GitHub users (due to the current marked dominance of GitHub). But it is still
not zero and might be about to increase again due to push back against
centralized services among FOSS developers.  In this paper we provide
methodological tools and improve upon the mitigation techniques proposed back
then. Instead of avoiding projects, one can start from cross-platform
datasets~\cite{swhcacm2018, swh-provenance-tr, swh-msr2019-dataset,
mockus2019woc} and measure the amount of shared VCS artifacts in the available
repositories.

 \section{Conclusion}
\label{sec:conclusion}

When relying only on forge-specific features and metadata to identify forked
repositories, empirical studies on software forks might incur into selection
and methodological biases. This is because repository forking can happen
exogenously to any specific code hosting platform and out of band, especially
when using distributed version control system (DVCS), which are currently very
popular among developers.

To mitigate these risks we proposed two different ways to identify software
forks solely based on intrinsic VCS data and development history: a
\emph{shared commit forks} (type 2) and a \emph{shared root directory forks}
(type 3) definition of software forks, as opposed to \emph{forge forks} (type
1) which are identifiable only when created on specific code hosting platforms
by, e.g., clicking on a ``fork'' UI element.  We also introduced the notions of
\emph{fork cliques} (set of repositories that share parts of a common
development history) and \emph{fork networks} (repositories linked together by
pairwise fork relationships) as ways to understand and quantify larger sets of
forks when using non-transitive definitions of forking.

Via empirical analysis of 40+\,M repositories using the GHTorrent and \SWH
datasets we quantified the amount of type 2 and type 3 forks that are not
recognizable as type 1 forks on GitHub, which appears to be substantial: +9\%
forks for type 2 forks, +37\% more for type 3.

We also showed that the aggregation/merge dynamics into larger clusters of
related repositories upon changing fork definitions is not just an absorption
phenomenon into a ``super attractor'' cluster, but that it concerns all
clusters: smaller ones are absorbed into larger ones of any size.

The methodological implications of our findings are that:
\begin{itemize}

\item Empirical software engineering studies on software forks aiming to be
  exhaustive in their coverage of forked repositories should consider using
  fork definitions based on \emph{shared VCS history} rather than trusting
  forge-specific metadata.

\item Depending on the research question at hand, the objects of studies to
  consider when looking at repositories involved in forks are either \emph{fork
    networks} or \emph{fork cliques}. The latter have the advantage of
  excluding cases that exist in the wild (e.g., on GitHub) in which
  repositories that do not share VCS artifacts might end up in the same fork
  network due to transitiveness.

\item Any set of repositories can be partitioned in accordance with its
  relevant shared commit fork cliques by computing its \emph{fork p-clique}
  partition function. This way of grouping together repositories that are all
  type 2 forks of each other is easily substitutable to partition approaches
  based on forge fork metadata.

\end{itemize}

\begin{acks}
  The authors would like to thank Théo Zimmermann for his careful review and
  comments on an early version of this paper.
\end{acks}

\clearpage

\end{document}